\documentclass[sigplan,10pt]{acmart}
\renewcommand\footnotetextcopyrightpermission[1]{}
\usepackage[english]{babel}
\usepackage{blindtext}

\usepackage{graphicx}
\usepackage[tt=false, type1=true]{libertine}
\usepackage[varqu]{zi4}
\usepackage[T1]{fontenc}
\usepackage{enumitem}

\usepackage{graphicx}
\usepackage{graphics}
\usepackage[utf8]{inputenc}
\usepackage{tabularx}
\usepackage{array, multirow} 
\usepackage{makecell}
\usepackage{enumitem}
\usepackage{array,multirow}
\usepackage[final]{listings}
\usepackage{color}
\usepackage{dblfloatfix} 
\usepackage{etoolbox}
\usepackage{float}
\usepackage{microtype}
\usepackage{subcaption}
\usepackage{filecontents}
\usepackage{multirow}
\usepackage{booktabs}
\usepackage{siunitx}
\usepackage{pifont}
\usepackage{xspace}
\usepackage{bm}
\usepackage{etoolbox}
\usepackage{pifont}
\usepackage{arydshln}
\usepackage{bm}
\usepackage{xurl}
\usepackage{subcaption}  

\PassOptionsToPackage{hyphens}{url}\usepackage{hyperref}

\usepackage{float}

\renewcommand\footnotetextcopyrightpermission[1]{} 
\setcopyright{none}

\acmDOI{}

\acmISBN{}

\acmPrice{}

\newcommand{\sys}{\textsc{NetReplica}\xspace}

\begin{document}

\settopmatter{printfolios=true}

\title{\sys: A Programmable Substrate for Bottleneck-Centric Network Data Generation}

\author{Jaber Daneshamooz$^*$,\hspace{0.2em}
Satyandra Guthula$^{*}$,\hspace{0.2em}
Jessica Nguyen$^{*}$,\hspace{0.2em}
William Chen$^\dagger$,\hspace{0.2em}
Sanjay Chandrasekaran$^{*}$,
Ankit Gupta$^\ddagger$,\hspace{0.2em}
Arpit Gupta$^{*}$,\hspace{0.2em}
Walter Willinger$^\mathsection$\\
University of California Santa Barbara$^*$ \hspace{0.3em} Northwestern University$^\mathsection$ \hspace{0.3em} Duke University$^\dagger$ \hspace{0.3em} AMD$^\ddagger$
}

\newcommand{\smartparagraph}[1]{\noindent{\bf #1}\ }
\newcommand{\att}{AT\&T\xspace}
\newcommand{\cv}{$cv$\xspace}
\newcommand{\metric}{Mbps/\$\xspace}
\newcommand{\diffcv}{$diff~cv_{max}_i$\xspace}
\newcommand{\maxcv}{$max~cv_{max}_i$\xspace}
\newcommand{\cmark}{\ding{51}\xspace}  
\newcommand{\eg}{e.g.\ }

\begin{abstract}
The behavior of Internet applications is shaped by congestion dynamics at bottleneck links, yet data capturing application behavior across diverse bottleneck regimes remains scarce. Bridging this gap requires a data-generation substrate that simultaneously provides \emph{controllability}, \emph{composability}, \emph{fidelity}, and \emph{replicability}—capabilities existing approaches struggle to achieve simultaneously. This paper introduces \sys, a programmable substrate for bottleneck-centric data generation guided by \emph{progressive disaggregation}: \sys (i) decouples {bottleneck intent} from {execution}, (ii) separates {static bottleneck attributes} from {dynamic congestion pressure}, and (iii) disaggregates observed demand dynamics from their original {trace context} via \emph{Cross-Traffic Profiles (CTPs)}. CTPs transform passive packet traces into reusable, composable pressure signals that can be selected and transformed to specify dynamic bottleneck behavior. Our evaluation shows that \sys satisfies the four requirements simultaneously and, in an ABR case study, generates datasets that remain realistic and expand coverage into underrepresented regimes, resulting in trained models that show improved performance (i.e., reducing transmission-time prediction error of the well-explored Fugu model by up to 47\%). Together, these results establish \sys as a practical and fully programmable bottleneck-centric data-generation substrate 
with direct value for developing production-ready protocols, applications, and learning artifacts.

\end{abstract}

\maketitle
\pagestyle{plain}
\settopmatter{printfolios=true}
\enlargethispage{-2\baselineskip} 
\begin{sloppypar}

\section{Introduction}
\label{sec:intro}
The behavior and performance of Internet applications—ranging from video streaming and conferencing to gaming and interactive cloud services—are governed by congestion dynamics at \emph{bottleneck links} along their end-to-end paths. While bottlenecks may arise at different locations, application outcomes are dominated by how capacity constraints, queueing, and competing traffic interact over time, producing bursty and intermittent congestion. These dynamics directly shape throughput, latency, loss, and ultimately user quality of experience (QoE). As a result, progress across networking research depends on access to data that captures how applications and protocols respond to \emph{diverse, time-varying bottleneck regimes}, rather than average or sanitized conditions. Specifically, many protocols and applications react directly to transient bottleneck signals such as queue buildup, loss, and delay variation; congestion-control mechanisms~\cite{schapira2019congestioncontrol, mlcc} and adaptive bitrate algorithms~\cite{pensive, swift} are therefore sensitive to the realism and coverage of the conditions under which they are evaluated. Likewise, for machine learning, training datasets must span a wide range of congestion behaviors, including rare slow paths and heterogeneous traffic mixes; otherwise models appear robust in controlled settings but fail when deployed in production environments~\cite{trustee, Meng2020}. Replicability and benchmarking pose related challenges: results are only comparable if experiments exercise similar \emph{bottleneck regimes} rather than merely similar code or configurations~\cite{bajpai2019dagstuhl}.

\smartparagraph{Why generating representative bottleneck-driven data is fundamentally hard.}
Generating data that faithfully captures application behavior across diverse bottleneck regimes is fundamentally difficult because bottleneck dynamics are simultaneously \emph{behavior-defining} and \emph{execution-dependent}. Application performance is shaped by fine-grained interactions among congestion control, queueing, and traffic demand at the bottleneck. Yet these interactions are only ever observed through specific executions—on particular paths, with particular link configurations, and under particular traffic mixes. As a result, the same bottleneck regime cannot be easily replicated, varied, or reused across environments.

This leads to a core abstraction gap in bottleneck-driven data generation. Researchers often reason about \emph{bottleneck regimes}—such as slow paths, bursty contention, or heterogeneous cross traffic—as abstract behaviors they wish to study. In practice, however, these regimes are never specified directly. Instead, they arise implicitly as side effects of where an experiment is run, how traffic is generated, or which traces happen to be collected. Without an abstraction that separates \emph{what bottleneck behavior is intended} from \emph{how and where it is realized}, bottleneck dynamics remain implicit, execution-bound, and difficult to control systematically.

\smartparagraph{Key requirements.}
Bridging this gap imposes a set of \emph{requirements} on any data-generation substrate intended to instantiate bottleneck-driven application behavior. 
For one, such a substrate must allow for the independent specification of static and dynamic attributes of the bottleneck link without entangling these choices with a particular path, trace, or execution environment (\emph{controllability}). At the same time, the substrate must permit independently specified attributes to be combined to construct more complex bottleneck regimes (\emph{composability}). Given a specification, the substrate must not only realize the intended network environment but also preserve closed-loop interactions among application logic, congestion control algorithms, and active queue management mechanisms under dynamic network conditions (\emph{fidelity}). Finally, these properties must hold consistently across repeated instantiations, requiring that the same specification can be \emph{re-instantiated} to induce statistically equivalent application behavior over time (\emph{replicability}).

\smartparagraph{Our approach.}
We address this challenge by introducing a data-generation substrate we call \sys that makes bottleneck dynamics explicit, reusable, and independent of execution context. The key insight behind \sys is that systematic bottleneck-driven data generation requires \emph{progressive disaggregation}: separating what bottleneck behavior is to be exercised from where it is realized, how it is parameterized, and how it was originally observed. Rather than embedding congestion dynamics implicitly in traces, paths, or infrastructures, \sys elevates bottleneck behavior itself into a first-class abstraction.

Concretely, \sys disaggregates bottleneck-driven data generation along three dimensions. First, it decouples {bottleneck intent} from execution, allowing the same bottleneck regimes to be instantiated across heterogeneous environments. Second, it separates {static bottleneck attributes}—such as capacity, base latency, buffering, and queue management—from {dynamic congestion pressure}, enabling systematic variation and targeted stress testing beyond static-only control. Third, it transforms passive packet traces into reusable \emph{Cross-Traffic Profiles (CTPs)}, thereby disaggregating observed demand dynamics from their original trace context.

CTPs capture how traffic demand evolves at a bottleneck---its intensity, burstiness, heterogeneity, and temporal structure---without binding these dynamics to the original path, applications, or users. \sys then programmatically selects and transforms CTPs to specify an intent's dynamic congestion pressure, while the application under study runs unmodified and remains closed-loop with the instantiated bottleneck. This preserves closed-loop realism where it matters—between the bottleneck and the application under study—while allowing demand dynamics to be reused, scaled, and composed across bottleneck configurations.

This progressive disaggregation enables \sys to satisfy the four key requirements concurrently. Concretely, disaggregation makes bottleneck intent independently specifiable (controllability) and enables independently specified components to be combined (composability). It preserves the closed-loop interactions that govern application behavior under time-varying conditions (fidelity), while allowing the same intent to be re-instantiated to induce statistically equivalent behavior over time (replicability). Together, these properties enable systematic coverage of bottleneck regimes for data generation. We detail in \S\ref{ssec:existing} how existing approaches fall short of providing this capability.

\smartparagraph{Contributions.}
This paper makes three key contributions:
\begin{itemize}[leftmargin=*,nosep]
  \item \textbf{A disaggregation-based design.}
  We employ progressive disaggregation---decoupling bottleneck intent from mechanisms, static bottleneck attributes from dynamic ones, and passive packet traces from their execution context---to satisfy the four key requirements concurrently.
  \item \textbf{Cross-Traffic Profiles (CTPs).}
  We introduce Cross-Traffic Profiles as first-class abstractions. They transform passive packet traces into composable, transformable representations of dynamic congestion pressure at bottlenecks while decoupling them from path-, user-, and environment-specific context.
  \item \textbf{\sys: system design and evaluation.} We design and implement \sys to instantiate the proposed progressive disaggregation. Our evaluation shows that \sys satisfies all key requirements. We present an illustrative case study (i.e., transmission-time prediction for ABR video streaming) to show the benefit of using \sys-generated data to train existing learning models that, rather than appearing robust in controlled settings, generalize to more challenging regimes that are typically underrepresented in existing training data.
\end{itemize}

\noindent
To be clear, \emph{this work is not about developing yet another ML model to improve ABR prediction inspired by~\cite{puffer}}. Instead, the paper (i) describes a programmable substrate for bottleneck-centric network data generation that is based on a progressive disaggregation architecture and introduces CTPs as first-class abstractions and (ii) shows that the proposed architecture delivers on what it promises. \sys has been open-sourced and is currently being used by several different research groups for use cases that require the generation of network traffic datasets that faithfully capture application behavior across diverse bottleneck regimes and include the study of TCP fairness, congestion control classification, the detection of network bottlenecks, the evaluation of network foundation models~\cite{demystifying}, and the development of network-security models. In short, 
\sys provides a principled and automated approach to generating high-quality datasets for studying Internet application behavior across diverse bottleneck regimes---a key requirement for speeding up networking research in general and in particular the development of generalizable or production-ready learning models.
\section{Background and Motivation}  
\label{sec:background}


\subsection{The Role of Bottleneck Links}
Internet applications communicate over end-to-end paths that traverse multiple network links, each contributing static path properties such as propagation delay and available capacity. Along such paths, one or more links may become congested due to limited capacity or contention from competing traffic. At any given time, application performance is typically governed by the \emph{active bottleneck link}, which constrains throughput and induces queueing delay and loss that drive transport- and application-level adaptation.

Note that while static properties—such as bottleneck capacity, queue size, and path latency—define the operating envelope of a connection, they do not uniquely determine application outcomes. For the same set of static attributes, different \emph{dynamic} conditions at the bottleneck---arising from cross-traffic intensity, burstiness, and its compositional heterogeneity---can induce markedly different congestion patterns and, in turn, different application behavior. Capturing these dynamics is therefore essential for understanding, evaluating, and reproducing how applications respond to network conditions. This observation motivates a bottleneck-centric view in which static bottleneck attributes and dynamic traffic variability are treated as distinct but interacting factors. 

\subsection{Key Data-Generation Requirements}
Since an application's performance is governed largely by the active bottleneck, and outcomes depend on the interaction between static bottleneck attributes and dynamic congestion pressure, a \emph{bottleneck regime} naturally comprises (i)~a static envelope and (ii)~a time-varying congestion-pressure process that drives contention within that envelope. The difficulty, however, is that data generation operates at the level of \emph{executions}. To generate data, a user must choose an \textit{execution context}---where the experiment runs, which hosts and paths participate, how competing demand is produced (traffic generators vs.\ traces), and which transport and queueing implementations and timing behaviors are exercised. These choices often couple static structure, dynamic pressure, and application behavior, making an intended regime difficult to express directly, vary along one dimension while holding others fixed, or re-instantiate consistently across environments.

These observations motivate four requirements for any bottleneck-centric data-generation substrate. It must (1)~let users specify static bottleneck attributes and dynamic congestion pressure independently, without those choices being dictated by the execution context (\emph{Controllability}); (2)~combine independently specified components predictably to construct richer regimes (\emph{Composability}); (3)~preserve closed-loop interactions among the application, congestion control, and queueing under time-varying conditions, while precisely realizing specified congestion signals over time (\emph{Fidelity}); and (4)~allow re-instantiation of the same regime specification to induce statistically equivalent application behavior over time, enabling meaningful comparison and benchmarking across runs and environments (\emph{Replicability}).

\subsection{Why Existing Approaches Fall Short?}
\label{ssec:existing}
Existing approaches typically satisfy some of the four requirements by construction, but not all of them concurrently.

\smartparagraph{Measurement datasets.}
Measurement-based datasets and platforms (e.g., Puffer~\cite{puffer} and M-Lab~\cite{mlab_speedtest}) capture end-to-end behavior in the wild, making them strong on {fidelity} to the observed execution contexts. However, bottleneck regimes are implicit in when and where measurements occur: users cannot specify a desired regime or systematically vary one dimension while holding others fixed. This limits {controllability} and {composability}, and it also limits {replicability} of a {desired} regime on demand.

\smartparagraph{Replay systems.}
Record-and-replay systems (e.g., Mahimahi~\cite{mahimahi}, CellSim~\cite{cellsim}, CellReplay~\cite{cellreplay}, Pantheon~\cite{pantheon}, iBox~\cite{ibox,ibox_short}, ndt-rr~\cite{ndt_rr}) rerun recorded executions, often preserving closed-loop effects and thus supporting {fidelity} to the recording. They also provide {replicability} for that specific execution. The limitation is that dynamic congestion pressure is tied to the trace and its collection context, making it difficult to independently control static and dynamic attributes or to combine demand components; thus {controllability} and {composability} are limited, and {replicability} does not readily extend to the same {intent} across heterogeneous environments. For a more detailed account of the limitations of each of these different replay systems, see Appendix~\ref{appendix:emulators}.

\smartparagraph{Programmable infrastructures.}
Programmable infrastructures and testbeds (e.g., FABRIC~\cite{fabric}, PINOT~\cite{pinot}, Netrics~\cite{netrics}, commercial clouds~\cite{aws2023overview,gcp2023overview}, and orchestration via NetUnicorn~\cite{netunicorn}) offer substantial control over the execution environment and can improve deployment-level {replicability}. However, the bottleneck regime is often inherited from the underlying fabric---bottlenecks may be absent, transient, or unrepresentative---so targeted regimes are difficult to realize without environment-specific engineering. As a result, these platforms provide weaker {controllability} and {composability} of bottleneck regimes as specifications, and can make it difficult to achieve high-{fidelity} realizations of intended regimes.


\smartparagraph{Takeaway.}
This prevailing gap motivates a substrate that treats bottleneck regimes as explicit specifications---with independently controllable static bottleneck attributes and dynamic congestion pressure---that remain composable, faithful under closed-loop execution, and re-instantiable over time.       

\section{Disaggregation-Driven Design}
\label{sec:design}

\subsection{Progressive Disaggregation}
\label{sec:progressive-disaggregation}
At its core, bottleneck-driven data generation requires more than realism or scale: it requires an \emph{intermediate representation}—an explicit, reusable specification of \emph{bottleneck regimes} (static bottleneck attributes plus dynamic congestion pressure). In today’s workflows, bottleneck regimes typically remain implicit: they become entangled with the execution context (where/how an experiment runs), conflate static and dynamic effects, or appear only through traces tied to the context under which they were collected. This coupling prevents existing workflows from meeting our four key requirements.

\smartparagraph{Learning from prior related structural obstacles.}
Similar structural obstacles have recurred across systems and have been addressed through principled \emph{disaggregation}: separating concerns that were previously entangled to enable portability, programmability, and reuse. Classic examples include protocol layering~\cite{chiang2007layering}, control--data plane separation in networking~\cite{openFlow,e2eSysDesign}, and the decomposition of intent, planning, and execution in operating systems and compilers~\cite{tanenbaum2015modern, aho2006compilers}. In each case, making intermediate structure explicit—rather than implicit in execution—enables systematic control.

Bottleneck-driven data generation exhibits the same failure mode. A bottleneck regime arises from fine-grained interactions among traffic demand, queueing, and congestion control, yet practitioners usually observe it only through particular executions—on specific paths, under specific link configurations, and with specific workloads. As a result, reproducing, varying, or reusing the same regime across different environments remains an unresolved problem. Treating bottleneck regimes as emergent side effects of traces, testbeds, or deployments resembles hard-coding execution logic without an intermediate representation: control remains indirect, replicability remains limited, and fidelity remains fragile.

\begin{figure}
    \centering
    \includegraphics[width=.8\linewidth]{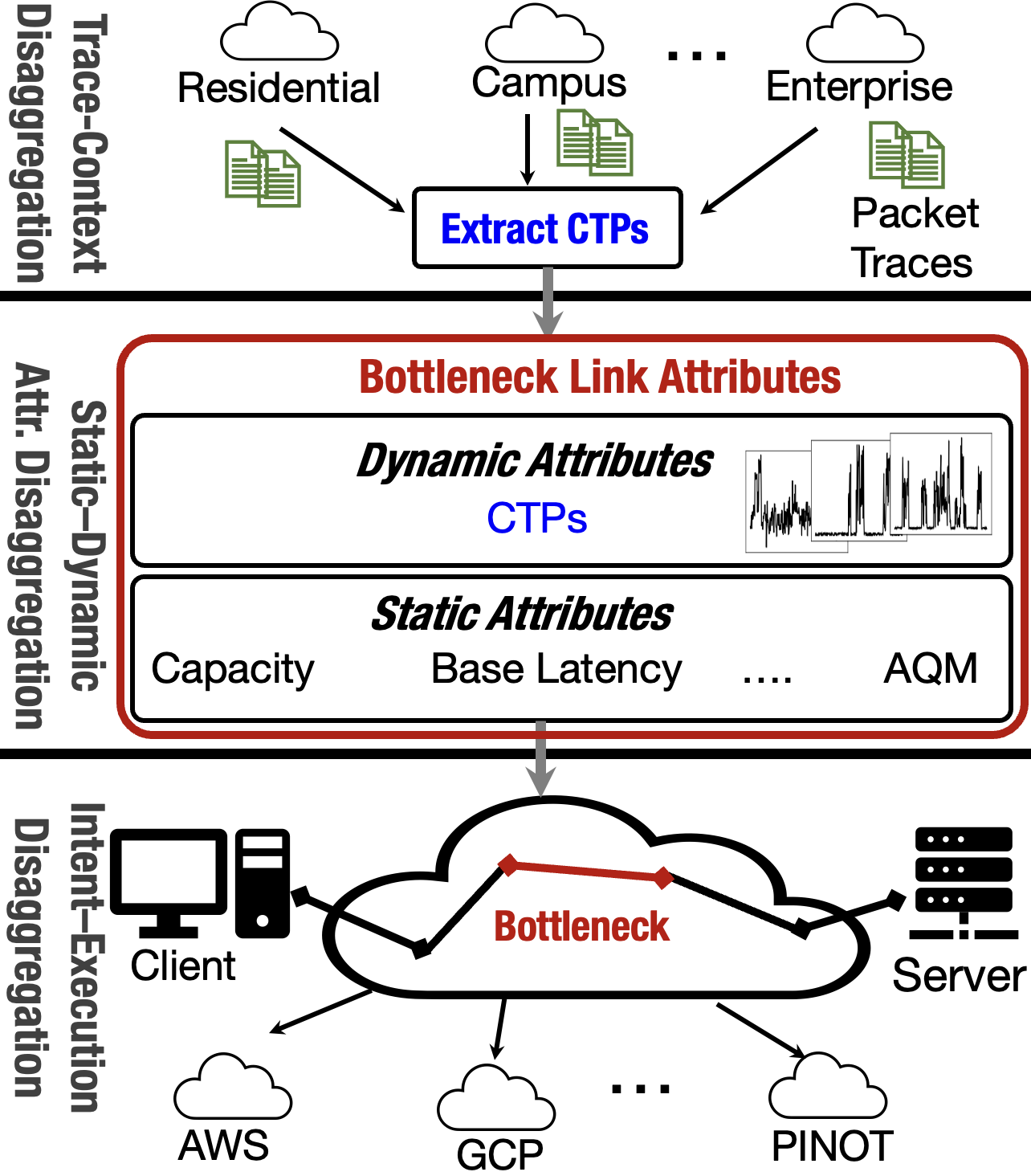}
    \vspace{-10pt}
    \caption{\sys overview: progressive disaggregation in action for bottleneck-centric data generation.}
    \label{fig:overview}
    \vspace{-15pt}
\end{figure}

\smartparagraph{Design principle overview.}
In response, \sys follows a unifying systems principle we call \emph{progressive disaggregation}. Rather than introducing a single abstraction boundary, \sys incrementally separates concerns that prior workflows conflate to enable bottleneck-centric data generation. Concretely, it disaggregates (i)~bottleneck intent from the \emph{execution context} in which experiments run (\S\ref{sec:intent-exec}), (ii)~static bottleneck attributes from dynamic congestion pressure (\S\ref{ssec:static-dynamic}), and (iii)~observed demand dynamics from their original \emph{trace context} via CTPs (\S\ref{sec:trace-context-disaggregation}). Figure~\ref{fig:overview} summarizes these separations as realized in \sys.

Although this design draws inspiration from thin-waist abstractions that decouple workload intent from execution in systems such as NetUnicorn~\cite{netunicorn}, it operates at a different layer: NetUnicorn separates \emph{where} workloads run from \emph{what} is executed, whereas \sys separates the \emph{bottleneck regime being exercised} from the contexts in which it is instantiated or observed.

\subsection{Intent--Execution Disaggregation}
\label{sec:intent-exec}
A fundamental obstacle to systematic bottleneck-driven data generation is the entanglement of \emph{what} bottleneck behavior is exercised with \emph{where and how} it is realized (the \emph{execution context}). In existing workflows, bottleneck behavior is rarely specified directly; instead, it emerges implicitly from the execution context; e.g., the deployment environment~\cite{netunicorn,pantheon}, the traffic source~\cite{mlab_speedtest,puffer,pensive}, or the traces being replayed~\cite{cellreplay,mahimahi}. As a result, the same intended bottleneck regime cannot be instantiated consistently across environments, or varied without changing the execution itself.

\smartparagraph{Proposed abstraction.}
Following the design principle of \emph{progressive disaggregation}, \sys separates \emph{bottleneck intent} from \emph{execution context}. Concretely, \sys introduces a first-class \emph{bottleneck-regime specification}: a declarative description of the congestion behavior the bottleneck should impose on the target application, independent of any particular testbed, cloud platform, deployment, or trace. Different execution contexts then realize the same specification without changing its meaning.

This separation enables (i)~\emph{intentional control}, because bottleneck regimes are specified directly rather than inferred from execution contexts, and (ii)~\emph{reuse}, because the same specification can be instantiated across heterogeneous execution contexts without redefining the experiment.

\subsection{Static--Dynamic Attribute Disaggregation}
\label{ssec:static-dynamic}
Intent--execution disaggregation makes bottleneck regimes explicit and portable, but it still treats a bottleneck regime as monolithic. In practice, a bottleneck’s impact on application behavior depends on more than static link configuration: capacity, base latency, buffering, and AQM define a structural envelope, yet the experienced regime depends critically on the time-varying contention injected into that envelope. Under the same static configuration, different arrival patterns induce qualitatively different regimes—smooth utilization, burst-driven queue buildup, persistent delay inflation, or intermittent loss—so static-only control collapses distinct behaviors and limits experimental precision and interpretability.

\smartparagraph{Proposed abstraction.}
Following \emph{progressive disaggregation}, \sys separates a bottleneck regime into two independently controllable components: \emph{static bottleneck attributes} and \emph{dynamic congestion pressure}. Static attributes specify the structural properties of the bottleneck (e.g., capacity, base latency, buffering/AQM), while dynamic congestion pressure specifies the temporal evolution of competing demand that drives contention against that structure.

This separation enables targeted stress testing (multiple dynamic pressure regimes against the same static bottleneck) and fair comparison (the same pressure applied across different static configurations). It also reduces misattribution by preventing execution artifacts from being mistaken for intrinsic bottleneck effects.

\subsection{Trace--Context Disaggregation}
\label{sec:trace-context-disaggregation}

Separating static structure from dynamic congestion pressure exposes a remaining challenge: \emph{how to represent dynamic congestion behavior itself in a form that is reusable, tunable, and independent of the trace context where it was originally observed}. A natural source of realistic congestion dynamics is packet-level traffic traces collected from production networks. Such traces encode rich temporal structure in traffic demand, but they do so only as \emph{context-bound artifacts}: each trace is inseparably tied to the specific path, link capacity, queueing configuration, application mix, and user behavior under which it was recorded. When replayed verbatim, these dynamics remain fixed to that trace context, preventing systematic reuse, controlled variation, or composition across environments. Thus, to treat dynamic congestion pressure as a controllable object, \sys disaggregates \emph{observed traffic dynamics} from the \emph{trace context} in which they were captured, while preserving the structure that drives realistic congestion behavior.

\smartparagraph{Proposed abstraction.}
\sys resolves this limitation by disaggregating observed traffic dynamics from their trace context through CTPs. A CTP is a reusable representation of \emph{dynamic congestion pressure} applied at a bottleneck.

\noindent
\textit{\underline{What do CTPs represent?}}
CTPs encode the \emph{temporal structure of aggregate demand} observed at a bottleneck; i.e., the pressure competing traffic applies to that bottleneck over time. Specifically, CTPs capture properties such as traffic \emph{intensity}, \emph{burstiness}, \emph{heterogeneity}, and coarse \emph{temporal correlations}—the aspects of demand that govern queue buildup, delay variation, and loss patterns. In doing so, CTPs describe how competing traffic arrives at the bottleneck and contends over time, without binding the profile to the particular end-host mechanisms that produced it.

This decoupling entails an explicit trade-off. CTPs omit per-flow semantics—application logic, transport feedback loops, user actions, and path-specific effects (RTT, downstream bottlenecks, etc.)—because these factors couple tightly to the trace context and change their meaning when the environment changes (e.g., TCP’s sending pattern depends on the RTT/loss it experiences). \sys accepts this loss because its goal is not to recreate the original execution, but to recreate the \emph{resulting bottleneck pressure signals} that drive the target application’s behavior. Preserving demand structure while shedding context-coupled causes yields a stable, reusable representation of bottleneck pressure that can be instantiated under different static link configurations and execution contexts.

\noindent
\textit{\underline{How are CTPs used?}}
CTPs act as \emph{exogenous, programmable pressure signals} at bottlenecks. During execution, background traffic follows the CTP and runs open-loop with respect to the target application, reproducing the prescribed demand dynamics independently of application behavior. This execution model is deliberate: reactive background traffic requires modeling the joint behavior of contributing applications, their transport protocols, and the users driving them, while maintaining a representative and stable mixture over time—an overly vague objective and unresolved problem.

\sys therefore adopts a \emph{hybrid replay model}: background congestion pressure is fixed by the CTP, while the \emph{target application remains fully reactive} to the resulting congestion signals. Closed-loop realism is thus preserved where it matters most—at the interaction between the bottleneck and the application under study—while background traffic remains controllable, reusable, and composable across experiments.

\smartparagraph{Operations over CTPs.}
Disaggregating traffic dynamics from their trace context introduces a design tension: demand structure must remain intact enough to reproduce realistic contention, yet flexible enough to support reuse across bottlenecks with different operating envelopes. \sys resolves this by defining CTPs as structured representations of congestion demand that retain the organization and variability of competing traffic while permitting controlled reinterpretation under new bottleneck constraints.

At this abstraction boundary, CTPs support three classes of transformation: (1)~\textit{selection} across regimes with comparable demand envelopes, (2)~\textit{adaptation} to accommodate mismatches between observed demand and target bottleneck capacity, and (3)~\textit{composition} of multiple demand sources into a single contention profile. These transformations preserve demand structure—rather than executions—making the trade-off between portability and fidelity explicit. By elevating this trade-off to the level of the representation, \sys enables congestion regimes to be reused and recombined intentionally, rather than emerging implicitly from trace provenance or execution context.

\smartparagraph{Consequences.}
Trace--context disaggregation elevates CTPs from fixed trace recordings to reusable specifications of dynamic congestion pressure. Rather than reproducing a particular trace context, \sys uses CTPs to preserve the distributional and temporal structure of demand that governs queueing, delay, and loss at bottlenecks, while allowing these dynamics to be instantiated across environments. In this setting, replicability means that the same CTP specification induces statistically comparable congestion behavior across runs, independent of where it is realized. By decoupling demand dynamics from trace provenance, this abstraction enables intentional reuse and composition of congestion regimes, completing the third step of progressive disaggregation.

\subsection{Progressive Disaggregation: Recap}
The proposed abstractions, guided by the principle of progressive disaggregation, map directly to our four requirements: \emph{controllability} (independent knobs for intent, static structure, and dynamic pressure), \emph{composability} (mix-and-match intent, structure, and pressure; select/adapt/compose pressure), \emph{replicability} (the same specifications re-instantiate comparable regimes across runs and environments), and \emph{fidelity} (preserve realistic queueing signals and closed-loop application--bottleneck interaction).

\section{\sys Implementation}
\label{sec:implementation}

\sys is implemented as a modular data-generation substrate organized into three logical planes. The \emph{intent plane} captures a user-specified bottleneck-regime specification independently of the execution context in which experiments run. The \emph{representation plane} constructs and manages reusable descriptions of dynamic congestion pressure in the form of CTPs derived from production traces and decoupled from their trace context. The \emph{execution plane} instantiates these specifications on concrete infrastructures, programs the bottleneck, and runs target applications unmodified.


\smartparagraph{Interfaces and mechanisms.}
At the intent plane, \sys provides \texttt{Link()} and \texttt{Bottleneck()} objects for specifying static bottleneck attributes---capacity, base latency, buffering, shaping, and queue management---without binding the specification to any particular execution context. Together, these objects define \emph{what} bottleneck behavior to exercise, independent of \emph{where} it is realized.

At the representation plane, \texttt{CrossTraffic()} objects manage reusable descriptions of dynamic congestion pressure derived from packet traces. Raw traces are processed into CTPs via \texttt{extract()}, indexed and retrieved by statistical attributes such as intensity, burstiness, and heterogeneity via \texttt{select()}, and applied at bottlenecks via \texttt{replay()}. \sys also supports adapting via \texttt{transform()} and composing multiple CTPs via \texttt{merge()} to control dynamic pressure systematically across static bottleneck envelopes.

\begin{figure}[t]
    \centering
    \begin{subfigure}{0.32\linewidth}
    \centering
    \includegraphics[width=\linewidth]{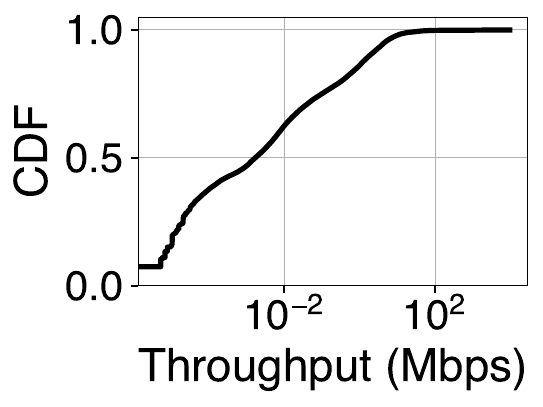}
    \vspace{-20pt}
    \caption{Intensity}
     \label{fig:ctp-intensity}
    \end{subfigure}
       \begin{subfigure}{0.32\linewidth}
    \centering
    \includegraphics[width=\linewidth]{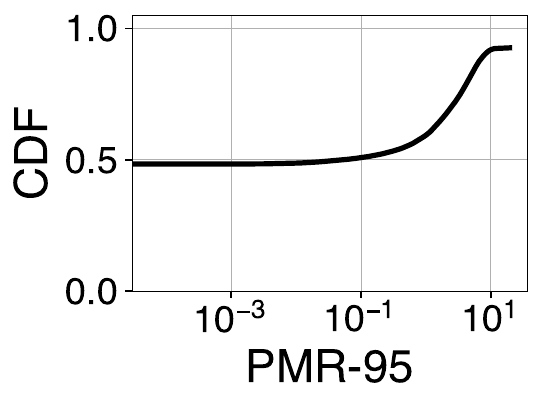}
    \vspace{-20pt}
    \caption{Burstiness}
     \label{fig:ctp-burst}
    \end{subfigure}
       \begin{subfigure}{0.32\linewidth}
    \centering
    \includegraphics[width=\linewidth]{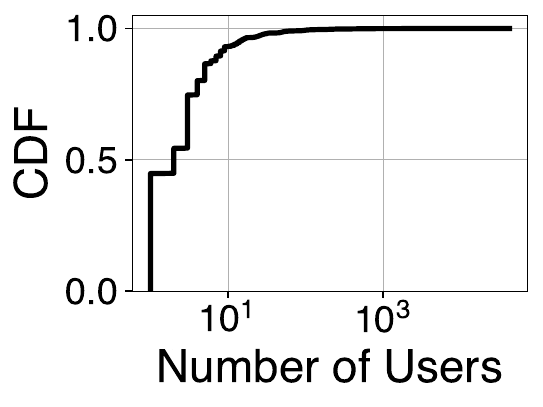}
    \vspace{-20pt}
    \caption{Heterogeneity}
     \label{fig:ctp-hetro}
    \end{subfigure}
    \caption{CTP characteristics from 15-minute gateway traces: (a)~intensity (mean throughput), (b)~burstiness (95th-percentile of peak-to-mean ratio of throughput as PMR-95), and (c)~heterogeneity (number of users).}
    \label{fig:ctp-chars}
\end{figure}

\begin{figure}[t]
    \centering

    \begin{subfigure}[b]{0.45\linewidth}
        \centering
        \includegraphics[width=\linewidth]{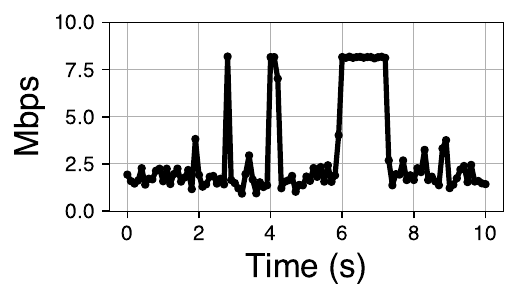}
        \vspace{-20pt}
        \caption{CTP1: $\downarrow$intensity,$\downarrow$bursty}
        \label{fig:ctp1}
    \end{subfigure}
    \hfill
    \begin{subfigure}[b]{0.45\linewidth}
        \centering
        \includegraphics[width=\linewidth]{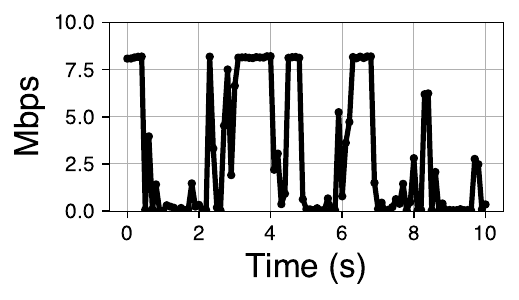}
        \vspace{-20pt}
        \caption{\textbf{CTP2}:$\downarrow$intensity,$\uparrow$bursty}
        \label{fig:ctp2}
    \end{subfigure}

    \begin{subfigure}[b]{0.45\linewidth}
        \centering
        \includegraphics[width=\linewidth]{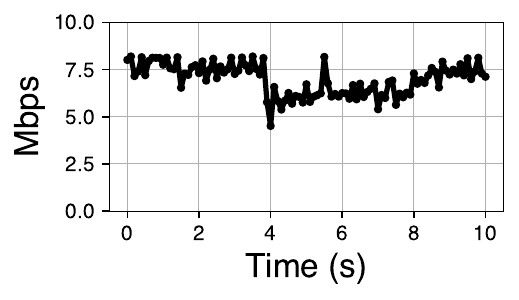}
        \vspace{-20pt}
        \caption{\textbf{CTP3}:$\uparrow$intensity,$\downarrow$bursty}
        \label{fig:ctp3}
    \end{subfigure}
    \hfill
    \begin{subfigure}[b]{0.45\linewidth}
        \centering
        \includegraphics[width=\linewidth]{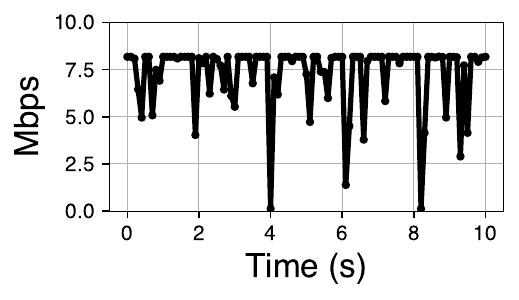}
        \vspace{-20pt}
        \caption{\textbf{CTP4}:$\uparrow$intensity,$\uparrow$bursty}
        \label{fig:ctp4}
    \end{subfigure}

    \caption{Visualization of throughput of four cross-traffic profiles with different intensity and burstiness.}
    \label{fig:ctp-attr}
    \vspace{-15pt}
\end{figure}

\begin{figure*}
    \centering
    \includegraphics[width=.9\linewidth]{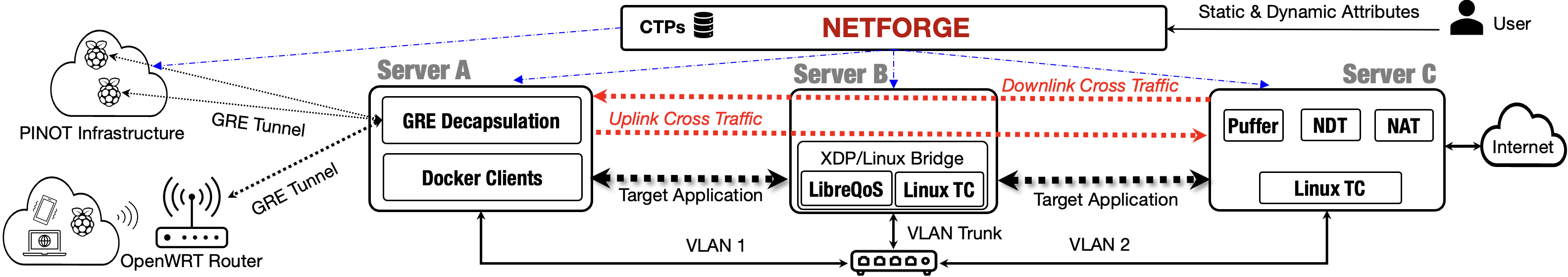}
\caption{Distributed \sys deployment: clients on Server~A, programmable bottleneck on Server~B, and endpoints on Server~C; specifications are compiled to backend configurations.}
\vspace{-10pt}
    \label{fig:implementation}
\end{figure*}

At the execution plane, \sys instantiates these specifications on concrete infrastructures using backend-specific mechanisms, including Linux traffic control and LibreQoS. Target applications run unmodified and interact with instantiated bottlenecks through standard network stacks. To support instantiation across environments, \sys additionally provides \texttt{NAT()} and \texttt{Tunnel()} objects for connecting experiments to the public Internet and constructing overlay topologies that span heterogeneous infrastructures.

\smartparagraph{Cross-traffic profiles.}
We next describe how \sys realizes trace--context disaggregation by constructing, indexing, adapting, and replaying CTPs extracted from packet traces.

\underline{\emph{Hierarchical demand representation.}}
Packet traces reflect contention at multiple spatial granularities (hosts, aggregates, subnets). \sys represents each CTP as a hierarchical demand object: it groups packets into bidirectional (anonymized) host-level contributors and aggregates them into prefix-based trees that preserve spatial locality. This representation captures heterogeneity, asymmetry, and traffic composition while remaining agnostic to application semantics.

\underline{\emph{Indexing and selection.}}
\sys indexes each CTP using statistical descriptors of congestion pressure. Temporal attributes include intensity, burstiness (e.g., Peak to Mean Ratio (PMR), Coefficient of Variation (CV)), and temporal correlation; structural attributes capture contributor composition (e.g., contributor count and upload--download asymmetry). \sys stores CTPs in a PostgreSQL database that supports multi-dimensional queries and retrieves them via the \texttt{select()} interface to target specific congestion regimes.

\underline{\emph{CTP corpus construction.}}
While \sys can ingest packet traces from any production network vantage point that observes bottleneck traffic and can convert them into CTPs, for our prototype, we benefit from access to our campus network and build the CTP corpus from a packet trace collected at a campus gateway (48k users, 8.2~Gbps peak) over a 15-minute interval using IRB-approved, privacy-preserving preprocessing. Our campus trace yields 1.4~M CTPs without windowing; using a 60\,s window with a 10\,s stride expands the corpus to 8.2~M CTPs, capturing transient and overlapping contention patterns.

Figure~\ref{fig:ctp-chars} shows that this single campus capture results in CTPs that span a wide range of intensity, burstiness, and heterogeneity and exhibit highly-variable behavior: many windows carry negligible load, while a smaller fraction contains strong and variable congestion pressure. After filtering out very low-intensity CTPs, we retain $\sim$230k CTPs---a small fraction of the total, but still a large design space for selection and transformation. Since CTPs represent congestion pressure independent of trace context, \sys reuses these dynamics across bottleneck configurations and execution contexts; the corpus requirement is diversity in pressure dynamics, not the trace location or duration. \sys indexes CTPs across these attributes and applies stratified sampling to expose both common and rare regions of this space. Figure~\ref{fig:ctp-attr} shows 10-second long segments of four representative CTPs (denoted CTP1, CTP2, CTP3, and CTP4), each drawn from a different strata.

\underline{\emph{Adaptation to target bottlenecks.}}
To instantiate CTPs under bottlenecks with different capacity constraints, \sys supports two adaptation modes realized through existing interfaces. \emph{Filtering} uses \texttt{select()} to choose CTPs whose offered load fits the target capacity envelope, while \emph{trimming} uses \texttt{transform()} to rescale amplitudes while preserving burst timing to respect shaping limits. \sys applies these choices explicitly at execution time.

\underline{\emph{Hybrid replay.}}
\sys drives background traffic from CTPs as open-loop cross traffic using \texttt{tcpreplay} with synchronized start, preserving inter-packet timing and burst structure. Linux \texttt{tc} or LibreQoS enforces shaping and queueing at the bottleneck, while target applications run unchanged and remain fully reactive to the resulting congestion signals.

\smartparagraph{Deployment.}
\sys supports multiple execution realizations of the same abstract bottleneck specification. Deployment changes only how \sys instantiates the specification on infrastructure; the intended bottleneck behavior and experiment semantics remain the same. Figure~\ref{fig:implementation} shows our prototype deployment.

In its \textit{distributed deployment}, \sys maps roles across multiple physical servers. Our prototype uses three servers (A, B, C), each equipped with dual 2.2~GHz Intel Xeon processors, 192~GB RAM, and dual Intel X710 10~Gbps NICs, interconnected via a Cisco SX550X 10~Gbps switch. Server~A hosts clients and target applications (e.g., Docker containers or Raspberry~Pis). Server~B hosts the programmable bottleneck and enforces shaping and queueing using Linux traffic control and LibreQoS (optionally with XDP). Server~C hosts application endpoints (e.g., Puffer and NDT) and auxiliary services (e.g., NAT). When experiments require connectivity beyond the testbed (e.g., to external infrastructures), \sys uses NAT and GRE~\cite{gre} tunnels. A compiler translates \texttt{Link}, \texttt{Bottleneck}, and \texttt{CrossTraffic} objects into backend-specific configurations across application hosts and bottleneck nodes. Unless specified otherwise, we use this distributed deployment for evaluations in this paper.

\sys also supports a \textit{standalone deployment} in which all components run on a single host using Linux namespaces, bridges, and traffic control. This realization executes the same specifications and APIs as the distributed deployment, enabling portability of bottleneck behavior across execution environments.

\section{System Evaluation}
\label{sec:syseval}
This section answers whether \sys satisfies all four key requirements---controllability (\S~\ref{ssec:controllability}), composability (\S~\ref{ssec:composability}), replicability (\S~\ref{ssec:replicability}), and fidelity (\S~\ref{ssec:fidelity})---that we want a programmable bottleneck-centric data-generation substrate to have. 

\vspace{-10pt}
\subsection{Experimental Setup}
\label{ssec:eval-setup}


\smartparagraph{Testbed and traces.} 
We run experiments on the three-server prototype shown in Figure~\ref{fig:implementation}. Servers A, B, and C each have dual 10~Gbps NICs, and Server~B provides the programmable bottleneck link. The topology has a single bottleneck and uses a FIFO queue unless explicitly varied; we illustrate \sys's ability to allow for bottleneck migration with a simple example in Appendix~\ref{appendix:migration}. As described in \S~\ref{sec:implementation}, we draw cross traffic from packet traces collected at one of our campus gateways (48k users, 8.2~Gbps bidirectional peak), preprocess it into millions of CTPs, and retain $\sim$230k of them after filtering out the very low-intensity CTPs. Unless otherwise stated, to illustrate some of our observations discussed below, we use the four representative CTPs shown in Figure~\ref{fig:ctp-attr}.


\smartparagraph{Configurations.}  
We evaluate \sys by progressively enriching the bottleneck link's specification through six configurations (see first 4 columns in Table~\ref{tab:configurations} in Appendix~\ref{appendix:configs}): \ding{182}~baseline forwarding through Server~B;  
\ding{183}~bottleneck at 3~Gbps with pFIFO~\cite{pfifo} queue;  
\ding{184}~stricter shaping at 10~Mbps;  
\ding{185}~addition of 100~ms base latency via {\tt netem};  
\ding{186}~introduction of bursty cross traffic replayed from CTPs with varying intensity and burstiness (Figure~\ref{fig:ctp-attr});  
\ding{187}~replacement of pFIFO with CoDel~\cite{codel} or FQ-CoDel~\cite{fqcodel}.  


\smartparagraph{Applications and logging.}  
We evaluate NDT speedtests over TCP Cubic (Linux default), with the server on Server~C and the client on Server~A. This setup minimizes application-layer complexity and isolates the evaluation to \sys's ability to programmatically control the bottleneck link. Each run lasts 10--15~s. We sample throughput (bytes transferred over time) every 100~ms and extract RTT from {\tt tcp-info} at the same granularity. We use RTT minus the minimum RTT as a proxy for queueing delay.

\subsection{Controllability}
\label{ssec:controllability}
Controllability requires that each static bottleneck attribute exposed by \sys can be set independently, such that modifying one attribute affects only its intended dimension of network behavior. Accordingly, we test the hypothesis that changing a single static bottleneck attribute—capacity or base RTT—while holding all others fixed induces a monotonic change in the corresponding observable (throughput or latency, respectively), without causing unintended changes in unrelated observables.

\begin{figure}[t]
    \centering
    \begin{subfigure}[t]{0.49\linewidth}
        \centering
        \includegraphics[width=\linewidth]{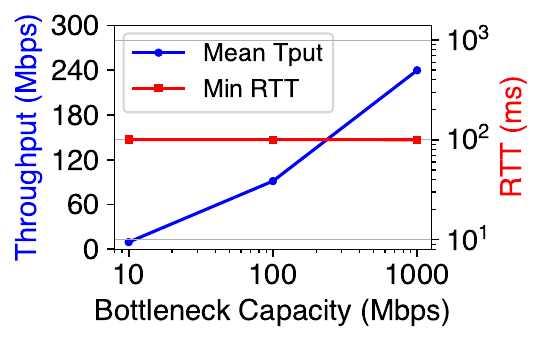}
        \vspace{-20pt}
        \caption{Capacity Sweep}
        \label{fig:controllability:left}
    \end{subfigure}\hfill
    \begin{subfigure}[t]{0.49\linewidth}
        \centering
        \includegraphics[width=\linewidth]{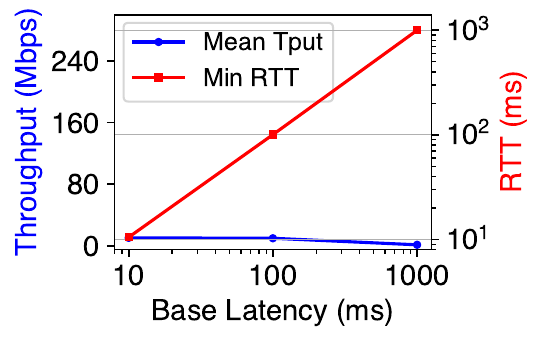}
        \vspace{-20pt}
        \caption{RTT Sweep}
        \label{fig:controllability:right}
    \end{subfigure}
    \vspace{-10pt}
    \caption{Independent control of bottleneck attributes: throughput scales with capacity, and RTT scales with latency.}
    \vspace{-10pt}
    \label{fig:controllability-static}
\end{figure}

To this end, all experiments use the single bottleneck topology, an unmodified NDT client and server, pFIFO queueing, and no cross-traffic. We perform two sweeps: (i) a capacity sweep, fixing base RTT at 100\,ms and varying bottleneck capacity among \{10, 100, 1000\}\,Mbps; and (ii) an RTT sweep, fixing bottleneck capacity at 10\,Mbps and varying base RTT among \{10, 100, 1000\}\,ms. 
Figure~\ref{fig:controllability-static} reports mean throughput and minimum RTT for the two cases, which capture the intended effects of capacity and latency, respectively. We observe that throughput scales monotonically with configured capacity while minimum RTT remains invariant across the capacity sweep, and vice versa. Table~\ref{tab:configurations} in Appendix~\ref{appendix:configs} offers a more detailed view of \sys's controllability across different configurations (\ding{182}–\ding{187}), reinforcing the controllability argument. Together, these results confirm that static bottleneck attributes in \sys are independently controllable, as required by the abstraction in \S\ref{ssec:static-dynamic}, and that adjusting one attribute does not implicitly alter others.

\begin{figure}[t]
    \centering
    \begin{subfigure}{0.23\textwidth}
        \centering
        \includegraphics[width=\linewidth]{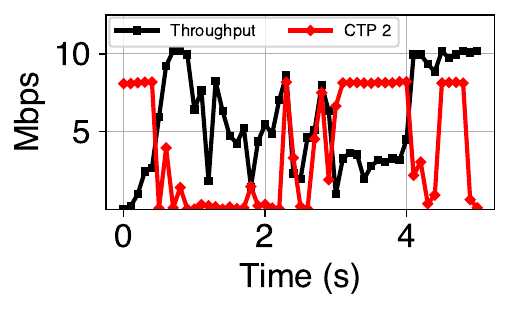}
        \vspace{-20pt}
        \caption{CTP~2 (Thput)}
        \label{fig:ctp2-thput}
    \end{subfigure}
    \hfill
    \begin{subfigure}{0.23\textwidth}
        \centering
        \includegraphics[width=\linewidth]{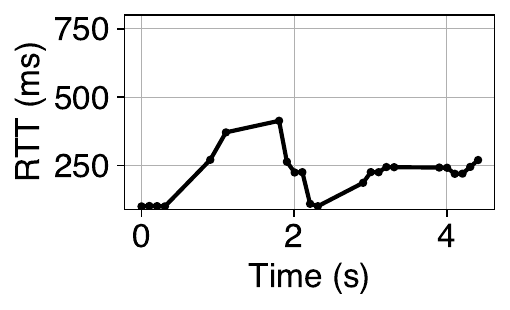}
        \vspace{-20pt}
        \caption{CTP~2 (RTT)}
        \label{fig:ctp2-rtt}
    \end{subfigure}
    \hfill
    \vspace{1ex}
    \begin{subfigure}{0.23\textwidth}
        \centering
        \includegraphics[width=\linewidth]{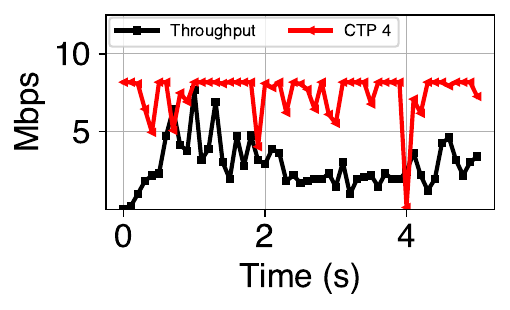}
        \vspace{-20pt}
        \caption{CTP~4 (Thput)}
        \label{fig:ctp4-thput}
    \end{subfigure}
    \hfill
    \begin{subfigure}{0.23\textwidth}
        \centering
        \includegraphics[width=\linewidth]{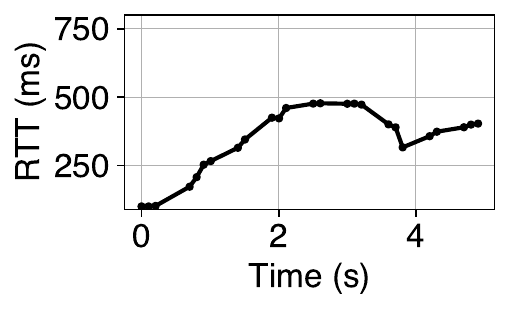}
        \vspace{-20pt}
        \caption{CTP~4 (RTT)}
        \label{fig:ctp4-rtt}
    \end{subfigure}
    \caption{Distinct dynamic attributes (CTPs) induce different RTT and throughput dynamics under identical static attributes, demonstrating composability. }
    \label{fig:dynamic_fidelity}
\end{figure}

\subsection{Composability}
\label{ssec:composability}

Composability requires that independently specified network attributes retain their semantic meaning when combined, such that their interactions reflect predictable network behavior rather than unintended interferences. Consequently, we test the hypothesis that when multiple bottleneck attributes---both static and dynamic---are specified together, the resulting behavior reflects known interactions among those attributes, rather than collapsing into opaque or unstable dynamics.

We evaluate composability along two complementary axes. First, we fix all static bottleneck attributes and vary only the dynamic congestion specification. Specifically, we use our single bottleneck topology with capacity fixed at 10\,Mbps, base RTT fixed at 100\,ms, FIFO queueing, an unmodified NDT client and server, and vary the CTPs (using CTP-2 and CTP-4, respectively). Second, we fix the dynamic congestion specification (CTP-2) and vary a static queueing attribute, comparing FIFO, CoDel, and FQ-CoDel under identical capacity and RTT. 



\begin{figure}[t]
    \centering
    \begin{subfigure}[t]{0.49\linewidth}
        \centering
        \includegraphics[width=\linewidth]{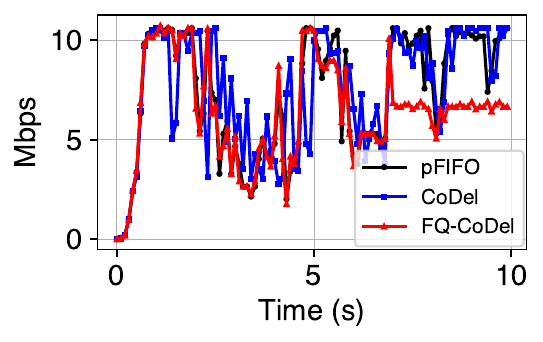}
        \vspace{-10pt}
        \caption{Throughput}
        \label{fig:aqms:left}
    \end{subfigure}\hfill
    \begin{subfigure}[t]{0.49\linewidth}
        \centering
        \includegraphics[width=\linewidth]{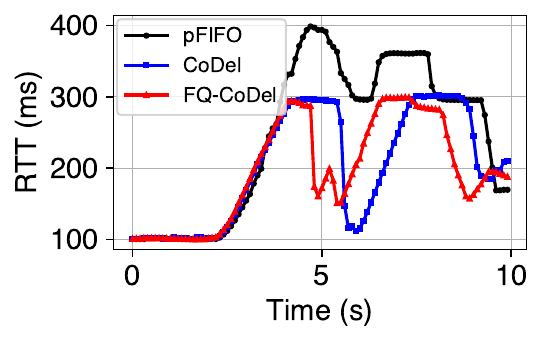}
        \vspace{-10pt}
        \caption{RTT}
        \label{fig:aqms:right}
    \end{subfigure}
    \caption{AQM $\times$ CTP-2 composability: fixing CTP-2 and varying the AQM (pFIFO/CoDel/FQ-CoDel) yields distinct (a)~throughput and (b)~RTT patterns.}
    \vspace{-12pt}
    \label{fig:aqm-composability}
\end{figure}

Figure~\ref{fig:dynamic_fidelity} shows that distinct CTPs induce clearly different RTT and throughput dynamics under identical static conditions, consistent with their specified burstiness and intensity. 
Figure~\ref{fig:aqm-composability} further shows that, under a fixed congestion profile, different AQMs exhibit characteristic queueing behavior (e.g., delay inflation under FIFO versus controlled delay under CoDel-based queues) without altering other bottleneck attributes. Since all non-varied attributes are held fixed in each experiment, these differences arise solely from the specified attribute under test, confirming that static and dynamic attributes compose predictably without interference. 

\subsection{Replicability}
\label{ssec:replicability}

Replicability requires that repeated instantiations of an identical bottleneck specification yield statistically similar behavior, such that observed differences can be attributed to specification changes rather than execution nondeterminism. Consequently, we test the hypothesis that executions generated from the same specification exhibit substantially lower variability than executions generated from different specifications. Table~\ref{tab:configurations} in Appendix~\ref{appendix:configs} summarizes this comparison using the 2-Wasserstein distance (2-WD)~\cite{WDSpringerNature, OptimalTransportWD}, which measures the normalized distance between different time-indexed network signals (i.e., RTT and throughput time series) and is sensitive to temporal misalignment. 

Note that we compute 2-WD over 100~ms RTT/throughput traces because replicability in our setting targets statistically equivalent closed-loop dynamics, not per-packet determinism. Packet-level matching would require sub-ms cross-host synchronization and tight control over OS and measurement timing; otherwise, inconsequential jitter would dominate distances. The bottleneck effects relevant to applications (queueing and rate dynamics) appear at 100~ms (base latency) timescales, so this granularity is sufficient for our replicability claim.

\begin{figure}[t]
    \centering
    \begin{subfigure}[t]{0.33\linewidth}
        \centering
        \includegraphics[width=\linewidth]{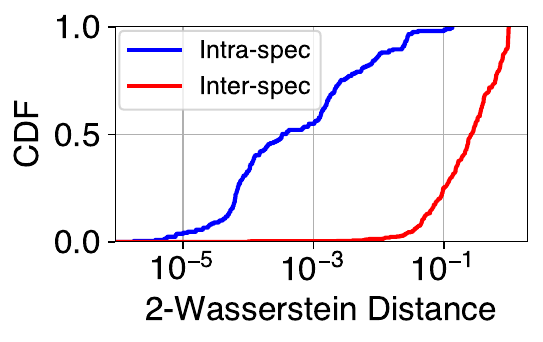}

        \caption{CDF}
        \label{fig:replicability-cdf}
    \end{subfigure}\hfill
    \begin{subfigure}[t]{0.33\linewidth}
        \centering
        \includegraphics[width=\linewidth]{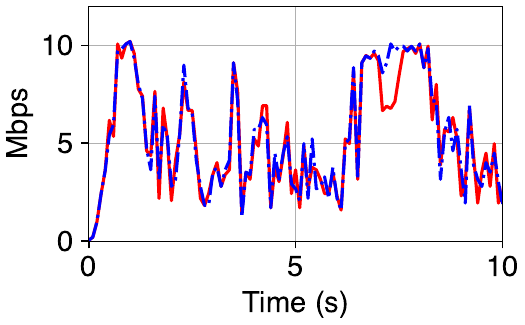}
        \caption{Throughput}
        \label{fig:similar_tp:left}
    \end{subfigure}\hfill
    \begin{subfigure}[t]{0.33\linewidth}
        \centering
        \includegraphics[width=\linewidth]{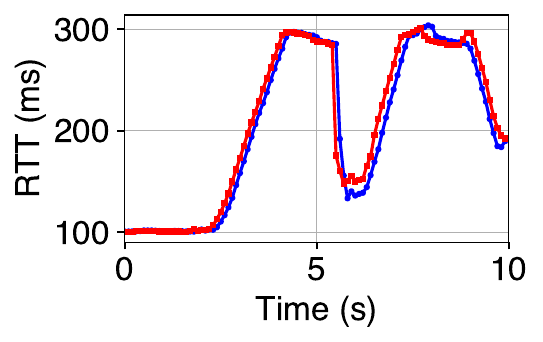}

        \caption{RTT}
        \label{fig:similar_tp:right}
    \end{subfigure}
\caption{Replicability under repeated instantiation. (a)~intra- vs.\ inter-specification 2-WD distance distributions; (b)~throughput and (c)~RTT traces for the worst-case intra-specification pair in configuration \ding{186}.}
    \label{fig:replicability-example}
\end{figure}

For each specification in Table~\ref{tab:configurations} in Appendix~\ref{appendix:configs}, we generate 10 independent instantiations and compute pairwise 2-WD distances between their RTT and throughput time-series. As a concrete illustration, we select \ding{186} and identify the two executions with the maximum intra-specification 2-WD distance. Figure~\ref{fig:replicability-example} plots the RTT and throughput time series for this worst-case pair, illustrating the largest temporal divergence observed among repeated instantiations of the same specification. Despite minor timing differences, both executions exhibit qualitatively similar congestion dynamics and operate under identical bottleneck conditions.

Figure~\ref{fig:replicability-cdf} shows the distributions of the 2-WD distance for throughput for intra-specification and inter-specification execution pairs. Across all configurations, the intra-specification 2-WD distribution is tightly concentrated near zero, while inter-specification distances are consistently larger, yielding a clear separation between identical and distinct specifications, confirming that observed variability is dominated by specification differences rather than execution instability.

\subsection{Fidelity}
\label{ssec:fidelity}

Fidelity requires that \sys preserves the semantic structure of user-specified static and dynamic attributes---especially dynamic congestion pressure---during execution, without collapsing, smoothing, or biasing temporal behavior. We therefore test the hypothesis that relative differences in dynamic congestion specifications persist after execution: specifications with higher temporal variability induce higher variability in observed network behavior under identical static conditions. 

Here we limit our evaluation to a qualitative illustration that uses the initial 5-second long window of each of the four representative CTPs shown in Figure~\ref{fig:ctp-attr}. In particular, the two plots on the left-hand-side of Figure~\ref{fig:dynamic_fidelity} present visual evidence of the impact of different dynamic network conditions on application behavior. While congestion specifications such as CTP2 and CTP4  with relatively high temporal variability induce relatively high variability in observed application behavior under identical static conditions (see Figure~\ref{fig:ctp2-thput} and Figure~\ref{fig:ctp4-thput}), the relatively low variability seen during the initial 5-second long windows of congestion pressure dynamics such as CTP1 and CTP3 also results in relatively low variability in observed application behavior (not shown).




For a more quantitative evaluation, we could fix all static bottleneck attributes, vary only the dynamic congestion pressure by replaying different CTPs, and use metrics such as the Coefficient of Variation (CV) to measure the variability of the CTP-offered load throughput (congestion pressure) and the observed throughput and RTT of the target application (i.e., NDT). However, CTPs with the same CV values can produce very different outcomes because intensity, temporal structure, and other CTP-specific features also matter. Unfortunately, we currently lack a CTP descriptor that overcomes CV's limited expressiveness and is able to specify CTPs more comprehensively. Recently considered embedding-based or representation learning approaches for time series data (e.g., Chronos~\cite{ansari2024chronos}, Chronos2~\cite{chronos2}, NetBurst~\cite{netBurst}) offer a potential path forward. They would enable a more fine-grained characterization of CTPs, with opportunities to represent, cluster, and index CTPs in ways that allow for their effective exploration, including the kind of quantitative analysis needed to establish that \sys satisfies the desired fidelity requirement.

\section{Data Evaluation}
\label{sec:dataeval}

This section uses ABR video streaming (Puffer)—with application-driven closed-loop behavior, unlike NDT speedtests—as a downstream case study to evaluate the realism (\S~\ref{ssec:realism}), coverage(\S~\ref{ssec:coverage}), and utility(\S~\ref{ssec:utility}) of \sys-generated \emph{ABR-specific} data for adapting existing learning models to more challenging, underexplored regimes. In particular, our interest here is not in building a better learning model but in considering domain adaptation — showing that \sys-generated training data increases network-condition diversity, which in turn improves the performance of existing models.

\subsection{Experimental Setup}
\label{ssec:setup}
\smartparagraph{Learning problem and model.}
We consider the transmission-time prediction problem for adaptive bitrate (ABR) streaming. Following prior work~\cite{puffer}, we use the Fugu model, which predicts the next chunk's transmission time from the past eight chunks (sizes and observed times) and TCP statistics---a task that directly dictates the decision-making by ABR algorithms~\cite{puffer, kairosVideoStreaming, a2br}.

\smartparagraph{Target domains.} We focus on challenging \emph{slow-path} conditions, where throughput is $\leq$6~Mbps and RTT $\leq$100--400~ms, consistent with prior definitions~\cite{puffer,pensive,yin2015control}. These regimes expose the core throughput--latency tradeoff, where accurate predictions most influence ABR decisions, yet they are relatively rare in the wild (see top row in Table~\ref{tab:dataset_size}) and thus important to emulate for bottleneck-centric data generation. We define four target domains with the same throughput $\leq$6~Mbps constraint and minimum RTTs of 100 to 400~ms.

\begin{table}[t]
\centering
\small
\caption{Puffer and \sys datasets for target domains with 6~Mbps thrpt cap and varying min RTTs.}
\vspace{-10pt}
\label{tab:dataset_size}
\resizebox{.85\columnwidth}{!}{
\begin{tabular}{lcccc}
& \multicolumn{4}{c}{\textbf{Target Domains}} \\
& {$\geq$100~ms} & {$\geq$200~ms} & {$\geq$300~ms} & {$\geq$400~ms} \\
\midrule
$PF^{new}$ & 440.1k & 126.5k & 33.4k & 10.8k\\
$NR$    & 651.5k & 261.5k & 120.3k & 48.6k \\
\end{tabular}
}
\vspace{-10pt}
\end{table}

\smartparagraph{Puffer datasets.}  
As a baseline, we use two weeks of production data from the Puffer service: September 17–30, 2022 ($PF^{old}$) and October 1–14, 2022 ($PF^{new}$). These periods coincide with when Fugu retraining had plateaued despite access to three years of prior logs. We filter $PF^{new}$ into four nested datasets, $PF_i$, according to the slow-path conditions described above. Each $PF_i$ contains only chunks with throughput $\leq$6~Mbps and minimum RTTs $\geq$ the target domain threshold, ensuring comparability across domains.

\smartparagraph{\sys datasets.}
We generate $NR_i$ datasets with \sys by shaping links to 4--10~Mbps and setting nine base latencies spanning the 10th--90th percentiles of $PF_1$ RTTs (76--444~ms)\footnote{These settings specify maximum link capacity and minimum latency; closed-loop dynamics typically yield lower application throughput and higher RTT.}. We sample 8k CTPs from the subset of campus traces with a mean throughput above 3~Mbps (half the target maximum). For video content, we use Puffer sample files from five TV channels~\cite{puffer_github}. Table~\ref{tab:dataset_size} shows that this approach yields $\sim$650~K data points, including $>4\times$ more samples in the more challenging $>400$~ms regime. 

\begin{figure}[thb]
    \centering
    \begin{subfigure}{0.49\linewidth}
    \centering
    \includegraphics[width=\linewidth]{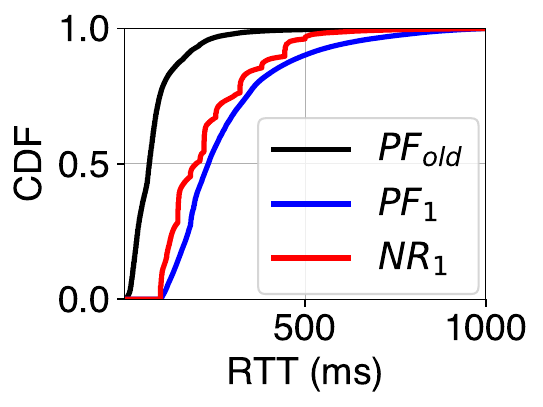}
    \caption{RTT}
     \label{md-rtt}
    \end{subfigure}
    \begin{subfigure}{0.49\linewidth}
    \centering
    \includegraphics[width=\linewidth]{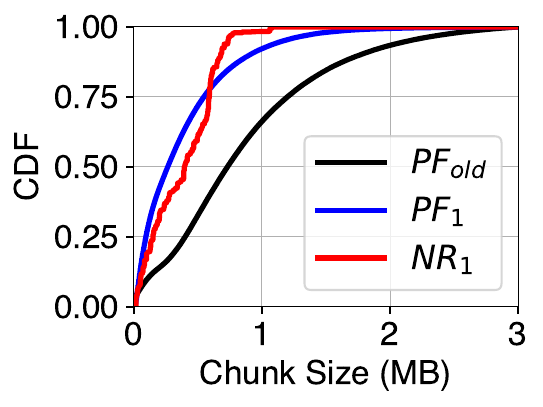}
    \caption{Chunk Size}
     \label{md-tt}
    \end{subfigure}
\vspace{-5pt}
\caption{Inherent specification bias in \sys-generated target-domain data.}
    \label{fig:stat_align}
    \vspace{-5pt}
\end{figure}

\smartparagraph{Inherent (unavoidable) specification bias.}
\sys-generated data can exhibit unavoidable specification bias because it is produced from a discrete set of static/dynamic intent settings and a finite video corpus. These design choices can skew feature distributions even when targeting the same domain. Figure~\ref{fig:stat_align} illustrates this effect: relative to the comparable Puffer target-domain slice $PF_1$, the \sys-generated target-domain data $NR_1$ shows shifted RTT and chunk-size distributions (e.g., RTT mass around discretized base-latency settings and chunk sizes reflecting the limited video set). In the remainder of this section, we evaluate whether \sys remains useful despite this inherent bias.

\subsection{Realism}
\label{ssec:realism}
To address concerns about the realism of \sys-generated datasets, we specifically test whether \sys produces \emph{plausible} executions (i.e., if \sys-generated samples lie within the empirical support of production executions (Puffer) without introducing unrealistic skews).


We represent each sample in $PF_1$ and $NR_1$ as an 8-chunk RTT/throughput trajectory. Before comparing trajectories, we normalize RTT and throughput by their standard deviations. For each $NR_1$ sample, we then compute its Euclidean distance to all $PF_1$ samples in the concatenated RTT+throughput trajectory space and record the nearest neighbors. Here, RTT and throughput serve as directly observed proxies for network conditions.

\begin{figure}[t]
    \centering
    \begin{subfigure}{0.49\linewidth}
        \centering
        \includegraphics[width=\linewidth]{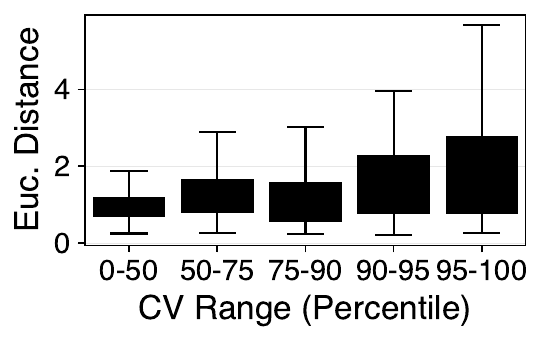}
        \caption{NN distances.}
        \label{fig:sigma_distance}
    \end{subfigure}
    \begin{subfigure}{0.49\linewidth}
        \centering
        \includegraphics[width=\linewidth]{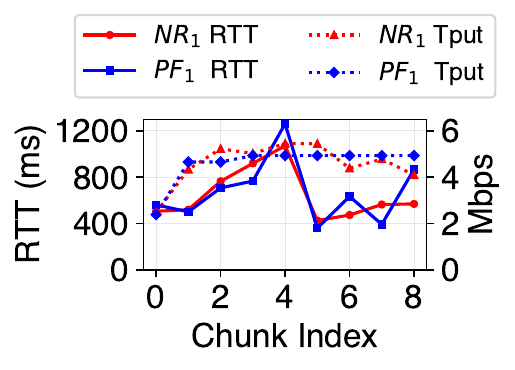}
        \caption{Example match.}
        \label{fig:sample1}
    \end{subfigure}
    \vspace{-5pt}
    \caption{\sys produces plausible executions. (a) Nearest Euclidean Neighbor distances from $NR_1$ to $PF_1$ across variability bins. (b) Example high-distance $NR_1$ sample and its nearest $PF_1$ neighbor.}
    \vspace{-15pt}
    \label{fig:combined_proximity}
    
\end{figure}

Note that we evaluate realism at chunk granularity because \sys targets chunk-level outcomes (as in Puffer). Packet-level matching is not available in Puffer logs and, even if it were, would be dominated by synchronization and host-level timing artifacts (e.g., OS scheduling and pacing) rather than bottleneck plausibility. 

Figure~\ref{fig:combined_proximity}(a) shows the distribution of Nearest Neighbor (NN) distances across five bins of a \emph{combined} variability score (percentiles) derived from RTT and throughput coefficient-of-variation values. Distances increase with variability: low-variability regimes are densely represented in $PF_1$, rare high-variability regimes have fewer close matches. To stress-test plausibility in the hardest regime, we examine a sample from the highest-variability bin (95--100th percentile) whose NN distance is 3.8 (90th percentile within that bin). Figure~\ref{fig:combined_proximity}(b) compares this $NR_1$ sample to its nearest $PF_1$ neighbor and shows qualitatively similar RTT and throughput trajectories. We observe similar qualitative agreement across 100 additional high-distance samples, indicating that \sys produces plausible executions even in rare, high-variability regimes.

\subsection{Coverage Expansion}
\label{ssec:coverage}

Next, we ask whether \sys provides \emph{novelty} by expanding effective coverage into bottleneck regimes that are empirically valid yet sparsely sampled in production. Production traces are inherently frequency-weighted: common regimes dominate the sample mass, while slow-path conditions (low throughput, high RTT) appear rarely. In contrast, \sys is intent-weighted and can deliberately densify these rare-but-valid regimes by rebinding dynamic congestion pressure (CTPs) to specified static attributes. Table~\ref{tab:dataset_size} quantifies this effect: under the same throughput constraint, \sys generates substantially more samples in the hardest regimes (e.g., more than four times for the $\geq 400$~ms regime), converting long-tail conditions into systematic training signal.

To examine where this additional coverage appears, we project the 62-dimensional feature space defined by Fugu's input data (i.e., a history of the past 8 chunks) into two dimensions with t-SNE and color each point by the original Fugu model’s prediction error (seconds). Figure~\ref{fig:tsne_projection} in Appendix~\ref{appendix:coverage} shows that $NR_{1}$ overlaps with $PF_{1}$ in dense regions—consistent with realism—while extending into nearby regions that Puffer covers sparsely. These added samples align with high-error areas, indicating that \sys expands coverage in precisely the regimes where the baseline model struggles most.

This expansion remains consistent with the nearest-neighbor realism in \S\ref{ssec:realism}: closeness in normalized RTT--Throughput space establishes plausibility but does not imply redundancy. Even within empirically valid neighborhoods, \sys generates distinct executions through targeted densification and varied interactions between dynamic pressure and the application across chunks. Overall, \sys expands datasets in a controlled and realistic way by filling long-tail gaps and targeting model blind spots, providing diversity for more robust and generalizable ML models.

\subsection{Adaptation to Slow-Path Regimes}
\label{ssec:utility}
Finally, we demonstrate the utility of \sys-generated data for improving transmission-time prediction (TTP) in slow-path regimes. We compare models trained or fine-tuned using domain-filtered production data ($PF_i$) against models trained or fine-tuned using \sys-generated data ($NR_i$). We evaluate on held-out $NR_i$ test sets because production slow-path logs are not sufficiently diverse within these extreme regimes to serve as a discriminative benchmark; $NR_i$ provides denser coverage of the rare, high-variability slow paths that dominate TTP errors.

\begin{figure}[t]
    \centering
        \includegraphics[width=\linewidth]{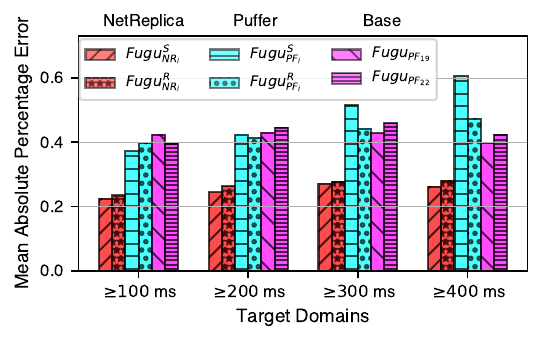}
         \vspace{-20pt}
    \caption{Performance of models trained with \sys vs. Puffer-generated datasets on slow-path regimes.}
    \label{fig:model-eval}
    \vspace{-20pt}
\end{figure}

We compare six Fugu variants. Two baselines, $\textit{Fugu}_{PF_{19}}$ and $\textit{Fugu}_{PF_{22}}$, are trained on Puffer data through Feb 2019 and Oct 2022, respectively. Domain-filtered Puffer baselines include Fugu$^{R}_{PF_i}$ (fine-tuning FuguPF22 on the training split of $PF_i$) and Fugu$^{S}_{PF_i}$ (training from scratch on $PF_i$). \sys-based models include Fugu$^{R}_{NR_i}$ (fine-tuning on $NR_i$) and Fugu$^{S}_{NR_i}$ (training from scratch on $NR_i$). We measure Mean Absolute Percentage Error (MAPE) on chunk transmission time. 
Figure~\ref{fig:model-eval} shows that \sys-trained models consistently reduce error on these densified slow-path stress tests, with up to 47\% lower MAPE in the hardest regimes, whereas training on domain-filtered Puffer subsets yields only modest gains. We note that \sys is intent-weighted and deliberately over-represents challenging, high-variability regimes; as a result, naive mixing or heavy fine-tuning on $NR_i$ can skew the effective training distribution and does not necessarily improve average performance on frequency-weighted production tests. Achieving uniform gains across both common and tail regimes requires principled sample selection and reweighting when combining generated and production data, which is outside the scope of this work.

\section{Looking ahead}
\label{outlook}

We view \sys as a promising initial step 
towards realizing a fully programmable bottleneck-centric data-generation substrate, with direct value for networking research efforts such as developing production-ready protocols, applications, and learning artifacts. While demonstrating feasibility, \sys's design also surfaces fundamental research challenges, some of which we discuss next.

\smartparagraph{Dynamic and heterogeneous bottlenecks.}
In this work, we use \sys to emulate a single fixed bottleneck per connection, which improves controllability but omits bottleneck migration (e.g., home WiFi $\rightarrow$ ISP access link). By composing two independently specified bottlenecks along a path, we illustrate in Appendix~\ref{appendix:migration} that \sys is able to support bottleneck migration. A related challenge is access heterogeneity: different media (e.g., cellular, DOCSIS, satellite) introduce distinct MAC and queueing dynamics. \sys currently composes WiFi with wired bottlenecks; extending support to additional access technologies broadens the space of realizable regimes but is left for future work.


\smartparagraph{Reactive vs non-reactive cross traffic.}
Currently, \sys replays passive traces as cross traffic. While this preserves realism in the sense that the target application remains fully reactive to congestion signals, it cannot adapt to network conditions or interact with target applications. This is a deliberate trade-off \sys makes: reactive replay would distort the burst structure and temporal correlations CTPs preserve. Real cross traffic responds at multiple layers—transport protocols adjust congestion windows under loss, and adaptive applications reduce bitrate under congestion-and what open-loop removes is a feedback loop no existing approaches are able to model faithfully. This includes state-of-the-art models like iBox~\cite{ibox} 
(see Appendix~\ref{appendix:emulators}). While generating cross traffic that is simultaneously realistic, diverse, and \textit{reactive} remains an unresolved problem,  NetGent~\cite{netgent} offers a potential path forward. It uses an AI-agent framework to translate natural-language workflow rules into executable user and application behaviors, enabling the generation of traffic that is repeatable, diverse across applications, and responsive to network conditions. Integrating such a framework with \sys could allow researchers to programmatically compose reactive cross traffic, capturing both transport-level adaptations and complex application-driven interactions.

\smartparagraph{Closed-loop data generation.}
\sys can instantiate a large space of bottleneck intents by combining static structure with CTPs; however, our current workflow leaves the choice of intents (and CTP transformations/compositions) to the user. A promising next step is to use downstream error or uncertainty to propose candidate intents and learn a generation policy under a fixed sampling budget (e.g., via bandits or reinforcement learning)~\cite{bandits,rl}. Such an approach would reframe synthetic data generation as \emph{learning which intents to instantiate} rather than \emph{deciding which particular packet traces to generate}. It would also mitigate a common failure mode of traffic synthesis that violates closed-loop transport/application semantics (e.g.,~\cite{feamster_traffic_synthesis, fanti_traffic_synthesis, apostolaki_traffic_synthesis}).



\section{Conclusion}
\label{sec:conclusion}
\sys makes bottleneck regimes a first-class, programmable object for data generation. Guided by \emph{progressive disaggregation}, \sys decouples bottleneck intent from execution context, separates static link structure from dynamic congestion pressure via Cross-Traffic Profiles (CTPs), and extracts reusable demand dynamics from packet traces independent of their trace context. Our evaluation shows that \sys satisfies controllability, composability, replicability, and fidelity, and that \sys-generated data remains realistic while expanding coverage into underrepresented regimes. In an ABR case study, model training with \sys-generated data reduces transmission-time prediction error by up to 47\% in challenging slow-path conditions, demonstrating the value of controlled, bottleneck-centric data generation.

\smartparagraph{Ethical considerations.}
We generate CTPs from campus traces collected using privacy-preserving methods~\cite{ontas}, and we do not collect or analyze personally identifiable information (PII).

\label{lastpage}

\end{sloppypar}

\bibliographystyle{ACM-Reference-Format}
\bibliography{ref,netunicorn}
\appendix

\newpage
\section{Challenges with Existing Emulators}
\label{appendix:emulators}

We expand on \S\ref{ssec:existing} by detailing why widely used emulators fail to
satisfy the four requirements concurrently, focusing on record-and-replay systems.
The common thread is that dynamic congestion pressure is either absent or bound to
a recorded trace, so these systems forgo the controllability, composability, and
dynamic fidelity that \sys provides.

\smartparagraph{Mahimahi, CellSim, and CellReplay.}
Mahimahi~\cite{mahimahi} replays captured traces alongside static attributes
(base latency, bandwidth), faithfully reproducing the recording but exposing no
control over dynamic cross traffic. CellSim~\cite{cellsim} adds traces for
cellular effects (interference, handovers) yet still does not model congestion
from competing traffic, and CellReplay~\cite{cellreplay} records separate light-
and heavy-workload traces to correct cellular bandwidth estimates. Crucially, all
three assume per-user queue isolation, so end-to-end delay is dominated by
self-interaction rather than contention at a shared bottleneck. Moreover, to
saturate highly variable cellular links and measure available bandwidth, these
tools rely on an UDP-based ``Saturator'' that maintains one-way delay between
roughly 750~ms and 3~s to approximate full link utilization. Such forced
saturation, however, can severely degrade the performance of other
applications---potentially halting services like video conferencing, which
cannot function with delays above 750~ms---thereby creating unrealistic
scenarios for cross-traffic evaluation. Since dynamic pressure cannot
be specified, varied, or composed and remains tied to the recording, these
systems lack \emph{controllability} and \emph{composability} over dynamic
attributes and provide only \emph{fidelity} to the original (cross-traffic-free)
recording.

\smartparagraph{Pantheon and iBox.}
Pantheon~\cite{pantheon} calibrates Mahimahi with five parameters (link rate,
propagation delay, DropTail threshold, i.i.d. loss, and an isochronous/Poisson
flag) to approximate diverse paths, judging success by matching average
throughput and 95th-percentile RTT within 17\%. Such coarse aggregates overlook
temporal protocol dynamics---similar percentiles can arise from very different
RTT patterns under bursty cross traffic---so realized \emph{fidelity} is
uncertain, and its multi-flow flooding offers no real control over cross traffic.
iBox~\cite{ibox,ibox_short} improves on this by explicitly modeling cross traffic,
but reduces it to one or a few TCP Cubic flows, discarding the heterogeneous flow
mixes, queueing dynamics, and application behaviors that \sys's CTPs preserve. It
further assumes a FIFO queue, estimates buffer size, propagation delay, and
bottleneck bandwidth via expensive Bayesian optimization rather than direct
specification, and captures only transport-level (not application-level)
reactivity. Both systems replay a limited set of recorded conditions, leaving
them weak on \emph{controllability}, \emph{composability}, and dynamic
\emph{fidelity}.

\begin{figure}[t]
\centering
\includegraphics[width=\linewidth]{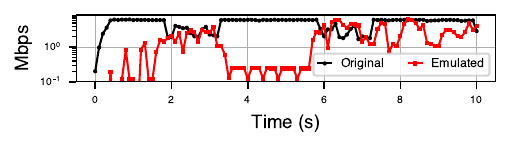}
\caption{ndt-rr's~\cite{ndt_rr} failure to distinguish self-induced from external congestion causes capacity underestimation and reduced emulation fidelity.}
\label{fig:netem-fidelity}
\vspace{-10pt}
\end{figure}

\smartparagraph{ndt-rr.}
ndt-rr~\cite{ndt_rr} reconfigures {\tt netem} from crowdsourced throughput, RTT,
and loss measurements, offering broad coverage of observed conditions. However, it
cannot distinguish external bottleneck congestion from an application's
self-induced congestion, misattributing application-driven fluctuations to the
network and inflating RTT even without cross traffic. To illustrate, we evaluate {ndt-rr} and select it over alternative tools like Mahimahi~\cite{mahimahi} because of its inherent controllability across billions of diverse measurements. To assess its realism, we used \sys to generate a ground-truth reference trace of the NDT speed test application under standard parameters (10~Mbps, 100~ms, pFIFO, no cross traffic), extracted network statistics using {ndt-rr}'s methodology, and generated {\tt netem} configurations at 100~ms granularity. Figure~\ref{fig:netem-fidelity} shows poor fidelity between the original and emulated throughput traces due to a fundamental limitation: {ndt-rr} cannot distinguish external bottleneck congestion from self-induced application congestion, and thus misattributes application-driven fluctuations to network dynamics. This issue persists at both coarse (1~s, as in the original work) and fine (20~ms, the highest tcp-info granularity available in M-Lab) configuration levels; even without cross traffic, {ndt-rr} interprets self-induced congestion as network variation and inflates RTT values. While coarse reconfiguration (1~s) fails to capture dynamic behavior, fine-grained updates (10–20~ms) introduce instability due to the system overhead of rapidly changing {\tt netem} parameters.
Discarding slow-start samples and redrawing
throughput from a fitted normal distribution further breaks the throughput--RTT dependency. Together, these artifacts undermine {ndt-rr}'s dynamic \emph{fidelity} and limit usable
dynamic \emph{controllability}.


\smartparagraph{Other emulators.}
Other tools~\cite{ns3, ns2, dummynet, mininet, mishra2021nemfi, ibox_short,
noble1997trace, mao2017neural, wang2025rattan, aireen2017wireless, chan2014opennet}
share these limitations. These studies either explicitly acknowledge their limitations or
follow the same patterns described above (e.g., the CellReplay analysis applies to
Rattan~\cite{wang2025rattan}). NetDiffusion~\cite{netdiffusion} instead synthesizes
protocol-constrained traces with diffusion models; though not directly comparable
to \sys, combining the two could broaden coverage with fewer experiments, provided
the realism of the generated data is validated using methodologies like the ones discussed in this paper.

\section{Bottleneck Migration}
\label{appendix:migration}

\begin{figure}[t]
    \centering
    \includegraphics[width=\linewidth]{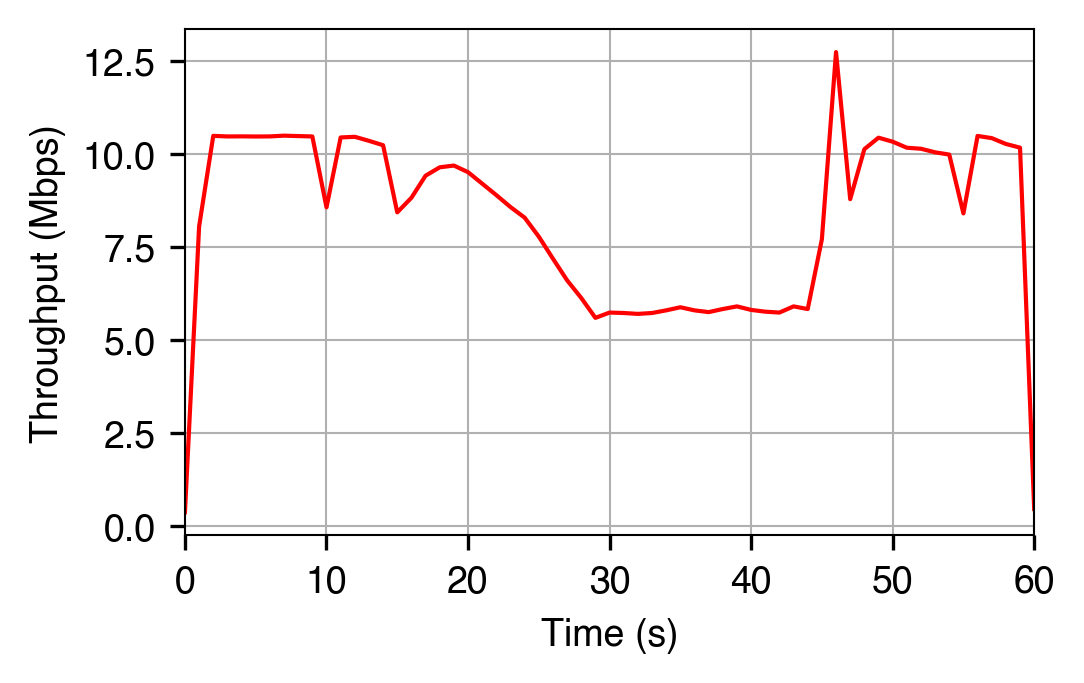}
    \caption{Bottleneck migration: Flow~A's throughput as its active bottleneck shifts from Link~1 to Link~2 once Flow~B begins contending on Link~2 at $t{=}15$\,s.}
    \label{fig:multibot}
\end{figure}

The experiments alluded to in \S\ref{sec:syseval} use a single fixed bottleneck per
connection. Here we illustrate that \sys's composability extends to the path
itself: by cascading two standalone \sys deployments (\S\ref{sec:implementation}),
each programming its own bottleneck link, \sys controls multiple bottleneck links
along a single path and can reproduce \emph{bottleneck migration}, where the
active bottleneck governing a flow shifts over time as competing demand changes
(\S\ref{ssec:static-dynamic}).

We chain two standalone \sys instances in series along the path of Flow~A, resulting in two independently specified bottleneck links, Link~1 and Link~2, shaped
at 10~Mbps and 15~Mbps, respectively. Flow~A downloads a large file for
60~seconds and traverses both links. In the absence of competing traffic, the
lower-capacity Link~1 is its active bottleneck. Fifteen seconds into the run, we
start Flow~B, which downloads a large file for 30~seconds and traverses Link~2
but not Link~1. As Flow~B ramps up, the residual capacity available to Flow~A on
Link~2 decreases; therefore the bottleneck for Flow~A
migrates from Link~1 to Link~2.

Figure~\ref{fig:multibot} plots Flow~A's throughput over the experiment's duration,
capturing this migration. Since each \sys instance is configured or emerges from
composition rather than bespoke configuration, this result demonstrates that
\sys realizes multi-bottleneck composable abstractions and that extending the abstractions we used beyond the single- and two-bottleneck scenarios is straightforward and simply requires cascading additional standalone \sys instances and placing more independently controlled bottleneck links along a path.

\section{Configurations and replicability results}
\label{appendix:configs}

For each configuration specified in Section~\ref{ssec:eval-setup}, Table~\ref{tab:configurations} reports for the target application (i.e., NDT) the median minimum RTT (column 5) and mean throughput (column 6) over 10 runs per configuration. The results demonstrate controllability across configurations, where different combinations of static and dynamic attributes in \sys produce distinct RTT and throughput outcomes. At the same time, columns 7 and 8 report the mean$\pm$std of the 2-Wasserstein distance between RTT and throughput, respectively, over 10 runs per configuration. The observed high consistency in the reported results demonstrates that \sys satisfies the replicability requirement.

\begin{table*}[t]
\centering
\caption{For the NDT speed test, we report the median minimum RTT and mean throughput over 10 runs per configuration (controllability/composability), and the mean$\pm$std 2-Wasserstein distance across all run pairs per configuration (replicability).}
\label{tab:configurations}
\resizebox{.9\textwidth}{!}{
\begin{tabular}{c|l|l|l|cc|cc}
{\textbf{Config}} & {\textbf{Capacity, Latency}} & {\textbf{Queue Policy}} & {\textbf{CTP}} & \textbf{Min RTT} & \textbf{Mean Throughput} & \textbf{RTT} & \textbf{Throughput} \\
\midrule
\ding{182} & - & pFIFO & & 0.33 & 1228.31 & 0.0077 ± 0.0019 &  0.0173 $\pm$ 0.0081 \\
\ding{183} & 3~Gbps & " & & 0.36 & 1325.21 & 0.0071 ± 0.0014 &  0.0146 $\pm$ 0.0075 \\
\ding{184} & 10~Mbps & " & & 0.56 & 10.21 & 0.0008 ± 0.0004 & 0.0047 $\pm$ 0.0104 \\
\ding{185} & 10~Mbps, 100~ms & " & & 100.66 & 9.64 & 0.0068 ± 0.0117 & 0.0001 $\pm$ 0.0000 \\
& 10~Mbps, 10~ms & " & & 10.55 & 10.18 & 0.0050 ± 0.0055 &  0.0001 $\pm$ 0.0000 \\
& 10~Mbps, 1000~ms & " & & 1000.68 & 1.16 & 0.0001 ± 0.0001 & 0.0001 $\pm$ 0.0000 \\
& 100~Mbps, 10~ms & " & & 10.58 & 101.21 & 0.0033 ± 0.0024 & 0.0000 $\pm$ 0.0000 \\   
& 100~Mbps, 100~ms & " & & 100.44 & 90.91 & 0.0055 ± 0.0057 & 0.0060 $\pm$ 0.0024 \\
& 100~Mbps, 1000~ms & " & & 1000.66 & 1.01 & 0.0001 ± 0.0002 & 0.0001 $\pm$ 0.0000 \\
& 1000~Mbps, 10~ms & " & & 10.40 & 939.65 & 0.0058 ± 0.0030 & 0.0115 $\pm$ 0.0314 \\
& 1000~Mbps, 100~ms & " & & 100.39 & 223.26 & 0.0000 ± 0.0000 & 0.0014 $\pm$ 0.0003 \\
& 1000~Mbps, 1000~ms & " & & 1000.84 & 1.03 & 0.0001 ± 0.0001 & 0.0001 $\pm$ 0.0001 \\
\ding{186} & 10~Mbps, 100~ms & " & CTP~1 & 100.51 & 7.06 & 0.0173 ± 0.0142 & 0.0016 $\pm$ 0.0004 \\
& " & " & CTP~2 & 100.69 & 7.05 & 0.0355 ± 0.0309 & 0.0221 $\pm$ 0.0097 \\
& " & " & CTP~3 & 100.52 & 4.76 & 0.0140 ± 0.0272 & 0.0026 $\pm$ 0.0016 \\
& " & " & CTP~4 & 100.50 & 4.88 & 0.0088 ± 0.0071 & 0.0011 $\pm$ 0.0005 \\
\ding{187} & " & CoDel & CTP~2 & 100.39 & 7.32 & 0.0192 ± 0.0113 & 0.0142 $\pm$ 0.0329 \\
& " & FQ-CoDel & CTP~2 & 100.48 & 6.72 & 0.0143 ± 0.0095 &  0.0272 $\pm$ 0.0419 \\
\end{tabular}
}
\end{table*}

\section{Visualizing Coverage}
\label{appendix:coverage}

\begin{figure}[H]
    \centering     
    \begin{subfigure}{\linewidth}
        \centering
        \includegraphics[width = .8\linewidth]{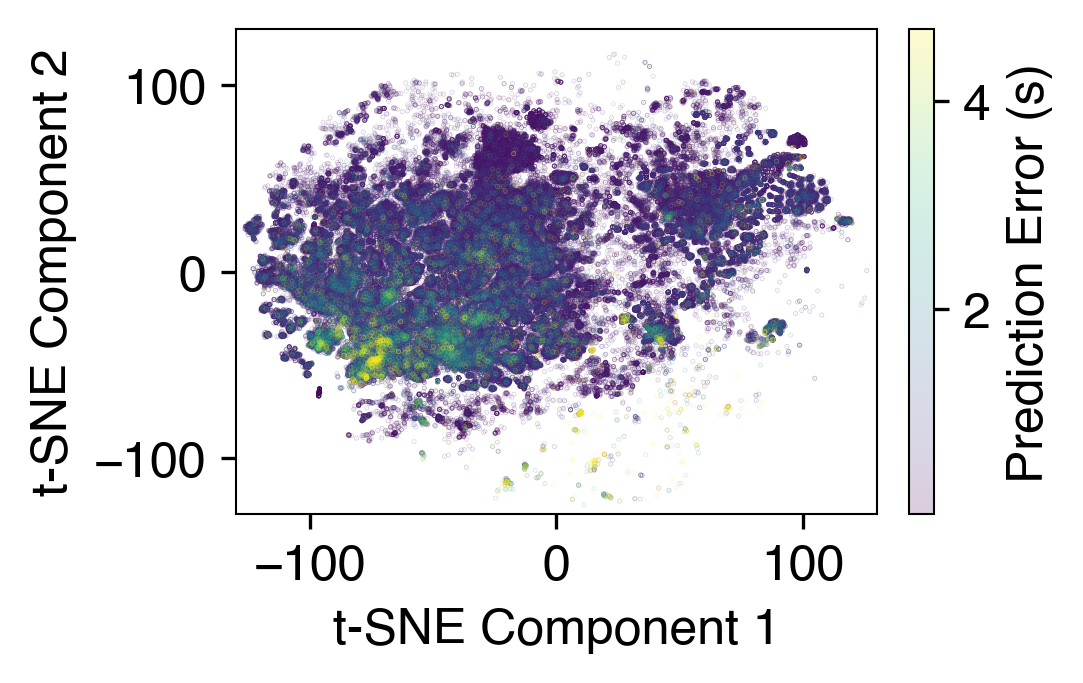}
        \caption{Puffer Dataset ($PF_1$)}
    \end{subfigure}

    \vspace{6pt}

    \begin{subfigure}{\linewidth}
        \centering
        \includegraphics[width = .8\linewidth]{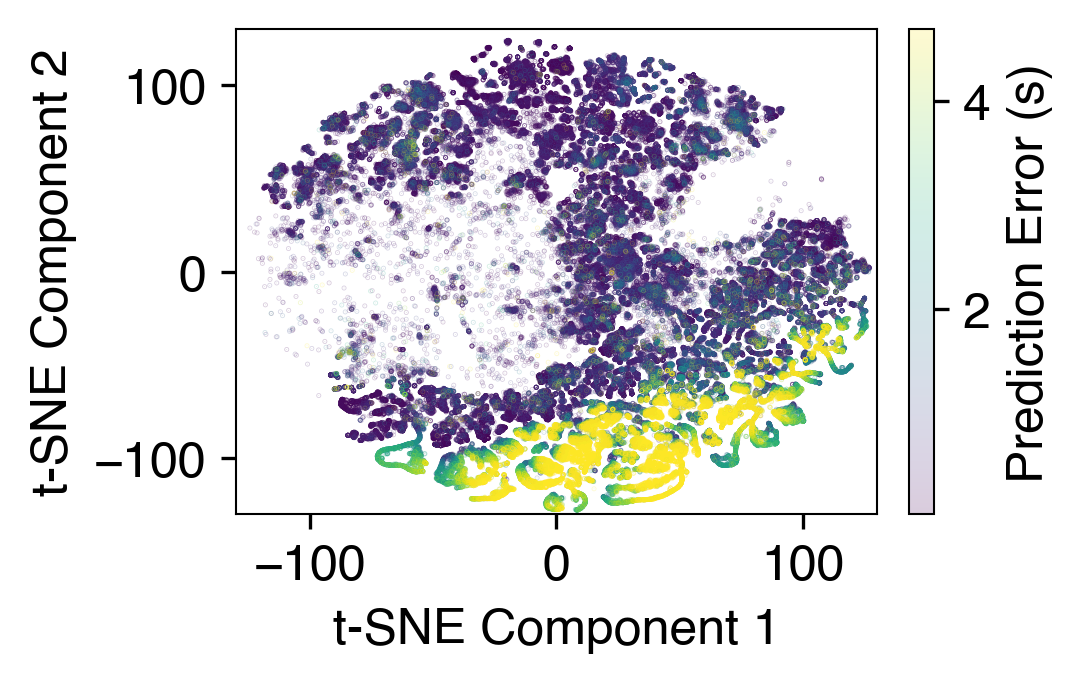}
        \caption{\sys Dataset ($NR_1$)}
    \end{subfigure}

    \caption{
    t-SNE visualization of 62-dimensional feature space.
    }

    \label{fig:tsne_projection}
\end{figure}

Figure~\ref{fig:tsne_projection} is a visualization in support of our argument in Section \S\ref{ssec:coverage}. Each point is colored by the original Fugu model’s prediction error (seconds). We observe that the \sys-generated dataset $NR_{1}$ (bottom) overlaps with the Puffer dataset $PF_{1}$ (top) in dense regions—consistent with realism—while extending into nearby regions that Puffer covers sparsely.

\end{document}